\documentclass[pra, twocolumn, floatfix]{revtex4}
\usepackage{graphicx}
\usepackage{amsmath, amsfonts, amssymb, bm}
\usepackage{braket}
\usepackage{color}
\usepackage[utf8]{inputenc}
\usepackage{subfigure}
\begin{document}
\title{ Formation of $\bar{\rm{H}}^+$ via 
electron-assisted three-body attachment of $\rm{e}^+$ to $\bar{\rm{H}}$  }
\author{ A. Jacob, C. M\"uller and A. B. Voitkiv}
\affiliation{ Institut f\"ur Theoretische Physik I, 
Heinrich Heine Universit\"at D\"usseldorf, 
\\ Universit\"atsstr. 1, 40225 D\"usseldorf, Germany }
\date{\today}
\begin{abstract} 

The formation of positive ions of antihydrogen $\bar{\rm{H}}^+$ via the three-body reaction (i) $\rm{e}^+ + \rm{e}^- + \bar{\rm{H}} \rightarrow \rm{e}^- + \bar{\rm{H}}^+$ is considered. In reaction (i), 
free positrons $\rm{e}^+$ are incident on 
antihydrogen $\bar{\rm{H}}$ embedded in 
a gas of low-energy ($\sim $ meV) electrons and,  
due to the positron-electron interaction, a positron is attached to $\bar{\rm{H}}$ whereas an electron carries away the energy excess. 
We compare reaction (i) with two radiative attachment mechanisms. One of them is (ii) spontaneous radiative attachment, in which the ion is formed due to spontaneous emission of a photon by a positron incident on $\bar{\rm{H}}$. The other is (iii) two-center dileptonic attachment which takes place in the presence of a neighboring atom B and in which an incident positron is attached to $\bar{\rm{H}}$ via resonant transfer of energy to B with its subsequent relaxation through spontaneous radiative decay. It is shown that reaction (i) can strongly dominate over mechanisms (ii) and (iii) for positron energies below 
$0.1$ eV. It is also shown that at the energies considered reaction (i) will not be influenced by annihilation and that the reaction $\rm{e}^+ + \rm{e}^+ + \bar{\rm{H}} \rightarrow \rm{e}^+ + \bar{\rm{H}}^+$ has a vanishingly small rate compared to reaction (i).  

\end{abstract}

\maketitle

\section{INTRODUCTION}

The reasons for the asymmetry between matter and antimatter in the universe are not yet (fully) understood which renders laboratory studies of antimatter very important. 

During the past few years, great progress has been made in the production of substantial amounts of the simplest atom of antimatter, the antihydrogen $\bar{\rm{H}}$ (see, e.g., \cite{Amoretti, Gabrielse_1, Enomoto, Maury, Andresen_1, Andresen_2, Gabrielse_2, Andresen_3}). Antihydrogen is (see, e.g., \cite{Amole_1, ALPHA, Amole_2}) and will be used for high-precision experiments and comparisons between the properties of antimatter and matter \cite{Holzscheiter}.

In this paper, we deal with the formation of the positive ion of antihydrogen $\bar{\rm{H}}^+$ which is the antimatter counterpart of the negative ion of hydrogen $\rm{H}^-$. It is of great interest as an intermediate particle in free fall experiments on the behavior of $\bar{\rm{H}}$ in the gravitational field of the Earth (GBAR experiment, see e.g., \cite{Perez, Cooke, GBAR} and references therein).

In a recent study \cite{Jacob_1}, we have focused on the formation of the $\bar{\rm{H}}^+$ ion via radiative attachment of a positron to $\bar{\rm{H}}$ in which the energy release is carried away by emission of a photon. Among the processes considered were spontaneous radiative attachment where $\bar{\rm{H}}^+$ is formed due to spontaneous emission of a photon by a positron incident on $\bar{\rm{H}}$ and two-center dileptonic attachment which becomes possible in the presence of a neighboring (matter) atom B and in which an incident positron attaches to $\bar{\rm{H}}$ via resonant transfer of excess energy to B driven by the two-center dileptonic interaction with subsequent stabilization of B via spontaneous radiative decay.

In this study, we consider the formation of $\bar{\rm{H}}^+$ via three-body attachment with an assisting electron, $\rm{e}^- + \rm{e}^+ + \bar{\rm{H}} \rightarrow \rm{e}^- + \bar{\rm{H}}^+$, in which positron capture by $\bar{\rm{H}}$ proceeds due to the positron-electron interaction where the energy excess is carried away by the electron. To our knowledge, this mechanism has been considered neither for the formation of $\bar{\rm{H}}^+$ nor for the closely related formation of $\rm{H}^-$ (via the reaction $\rm{e}^- + \rm{e}^+ + \rm{H} \rightarrow \rm{e}^+ + \rm{H}^-$). 

Note that the formation of $\bar{\rm{H}}^+$ ions can also occur in collisions between (excited) positronium $\rm{Ps}$ and $\bar{\rm{H}}$ via the capture reaction $\rm{Ps}+\bar{\rm{H}} \rightarrow \rm{e}^- + \bar{\rm{H}}^+$ (see, e.g., \cite{Cooke, Comini} and references therein).

The main goal of this paper is to find out whether three-body attachment with an electron (mechanism (i)) can be more efficient than the spontaneous radiative attachment (mechanism (ii)) and the two-center dileptonic attachment (mechanism (iii)). As will be shown, for low-energy electrons and positrons, mechanism (i) can strongly dominate over mechanisms (ii) and (iii) and at such energies the corresponding three-body reaction  
$ \rm{e}^+ + \rm{e}^+ + \bar{\rm{H}} \rightarrow \rm{e}^+ + \bar{\rm{H}}^+$ has vanishingly small rates. 

The paper is organized as follows. In Section II we obtain the $\bar{\rm{H}}^+$ formation rate for three-body attachment with an  electron. Section III contains numerical results and a comparative discussion of the three-body attachment and radiative attachment mechanisms. We summarize our main findings in Sec. IV.

Atomic units ($\hbar = |e| = m_e = 1$) are used throughout unless otherwise stated. 

\section{Theoretical consideration }

\subsection{ Three-body attachment with an assisting electron (3BAe)}

\begin{figure}[h!]
\centering
\includegraphics[width=8.5cm]{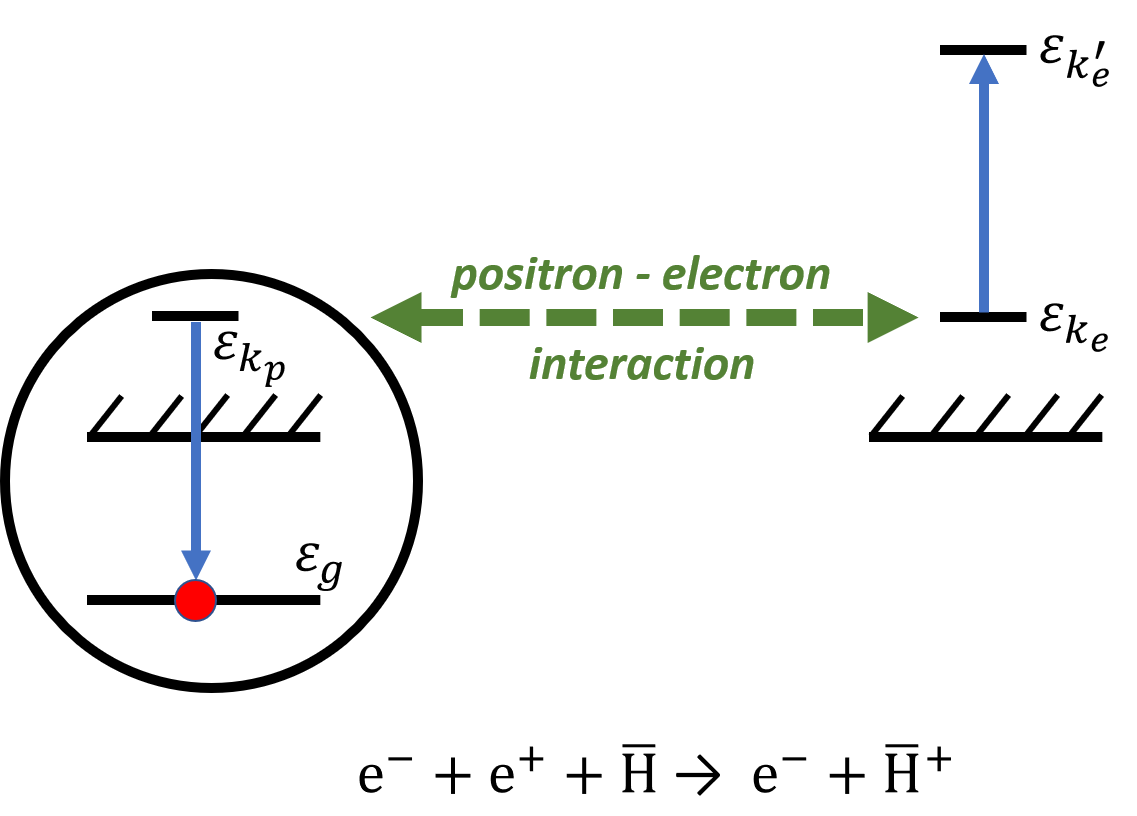}
\caption{ Scheme of three-body attachment with an assisting electron (3BAe).}
\label{fig:figure1}
\end{figure}

Let us consider an enviroment where free positrons and electrons move in a close vicinity of antihydrogen $\bar{\rm{H}}$. In this case, the positron-electron interaction may lead to attachment of a positron to $\bar{\rm{H}}$ where the released energy is carried away by an electron. A scheme of three-body attachment with an assisting electron is shown in Fig. \ref{fig:figure1}. Here, $\varepsilon_{k_p}$ ($\varepsilon_{g}$) is the energy of the incident (captured) positron and  $\varepsilon_{k_e}$ ($\varepsilon_{k_e'}$) is the energy of the incident (outgoing) electron.

We choose a reference frame in which $\bar{\rm{H}}$ is at rest and take the position of the antiproton as the origin. 
We begin our consideration of the process with the exact transition amplitude 
in the post form which reads (see e.g. \cite{post-form})  
\begin{eqnarray}
a_{3BAe} &=& -i \int_{-\infty}^{\infty} dt \,   \bigg \langle \bigg( \hat{H} - i \frac{\partial}{\partial t} \bigg)  \chi_f(t) \bigg|  \Psi^{(+)}_i(t) \bigg \rangle. 
\label{amplitude_exact}
\end{eqnarray}  
Here, $\Psi^{(+)}_i(t)$ is the exact solution 
of the full Schr\"odinger equation 
(whose Hamiltonian $\hat{H}$
includes all the interactions)    
which satisfies the 'incoming'  
boundary condition. 
Further, $\chi_f(t)$ is the final asymptotic state, 
which is a solution of the Schr\"odinger equation with 
the corresponding asymptotic Hamiltonian $\hat{H}_f $. 

The process under consideration 
is a four-body problem which is quite difficult to handle. 
Therefore, we shall treat this process 
(in approximate manner) 
as an effectively three-body problem in which the antiproton and the initially bound positron 
are considered as a single body which has a rigid structure and produces a short range potential.  

Taking into account that the Coulomb interaction between the incident electron and positron is much stronger than the interactions of these particles with the (neutral) antihydrogen, we approximate the exact state $\Psi_i^{(+)}(t)$ by 
\begin{eqnarray}
\Psi_i^{(+)}(\bm{r}_e, \bm{r}_p, t) &=& e^{i \bm{ P } \cdot {\bm R} } \, \, \psi_{\bm{\kappa}}^{(+)} ( \bm{ r }) 
\, \, e^{-i (\varepsilon_{k_p}+\varepsilon_{k_e}) t}. 
\label{initial_state_1}  
\end{eqnarray} 
Here, $\bm r_p$ ($\bm r_e$) is the space coordinate of 
the incident positron (electron),  
$ {\bm R} = (\bm{r}_e + \bm{r}_p)/2$ is the coordinate of the center-of-mass of the incident electron-positron pair,  and 
$\bm{ P } = \bm{k}_e + \bm{k}_p $ is  
their total momentum, where $\bm{ k }_p$ ($\bm{ k }_e$) 
is the asymptotic momentum of the incident positron (electron) with corresponding kinetic energy 
$\varepsilon_{k_p} = \bm{k}_p^2/2$ 
($\varepsilon_{k_e} = \bm{ k }_e^2/2$).  
Further, 
\begin{eqnarray}
\psi_{\bm{\kappa}}^{(+)} ( {\bm r} )   &=& \frac{1}{\sqrt{V_p V_e}} e^{\frac{\pi}{2 \kappa}} \Gamma \bigg(1-\frac{i}{\kappa}\bigg) e^{i \bm{\kappa} \cdot \bm{ r } } \nonumber \\
&& \times ~ F \bigg(\frac{i}{\kappa}, 1, i (\kappa r - \bm{\kappa} \cdot \bm{ r } ) \bigg) 
\end{eqnarray}
is the Coulomb wave function describing the relative motion 
of these two particles, where $\bm{ r } = \bm{r}_e - \bm{r}_p$ is the relative coordinate of the pair, 
${\bm \kappa} = ({\bm k}_e - {\bm k}_p)/2$, $V_p$ ($V_e$) is the positron (electron) normalization volume and $F(a,b,z)$ is the confluent hypergeometric function \cite{Abramowitz}.  

Thus, the state (\ref{initial_state_1}) describes 
the motion of two incident particles, the positron and electron, by fully accounting for the long-range Coulomb interaction between them but neglecting their interaction with the neutral antihydrogen.   

The final state $\chi_f(t)$ is chosen as 
\begin{eqnarray}
\chi_f(t) = \phi_b(\bm{r}_p) \, 
\varphi_f(\bm{r}_e, \bm{r}_p) \,  e^{-i (\varepsilon_{g}+\varepsilon_{k_e'}) t} ,     
\label{final_state}
\end{eqnarray}  
where $ \phi_b(\bm{r}_p)$ is the (undistorted) bound state of the positron, which was attached to the antihydrogen,    
and $\varphi_f(\bm{r}_e, \bm{r}_p)$ the state describing 
the outgoing electron moving in the field of the $\bar{\rm{H}}^+$ ion formed.   
Regarding the $\bar{\rm{H}}^+$ ion as an effectively 
one-positron system, in which the positron attached is a weakly 
bound "outer" positron moving in the short-range field of the ionic core, we approximate the state $\phi_b$ by
\begin{eqnarray}
\phi_b (\bm{r}_p) &=& N \,  \frac{e^{-\alpha r_p} - e^{-\beta r_p}}{r_p}.
\label{ion_bound_state_p}
\end{eqnarray}
Here, $N = \sqrt{\frac{\alpha \beta (\alpha+\beta)}{2 \pi (\beta-\alpha)^2}}$, $\alpha = 0.235$ a.u. ($\varepsilon_g = \alpha^2/2 = 0.0275$ a.u. $\approx 0.748$ eV is the binding energy) and $\beta = 0.913$ a.u.. The wave function \eqref{ion_bound_state_p} was derived by using a nonlocal separable potential of Yamaguchi \cite{Yamaguchi} for describing a short-range effective interaction of the “active” positron with the core of $\bar{\rm{H}}^+$. 

We choose 
$\varphi_f$ to be given by a Coulomb wave 
\begin{eqnarray}
\varphi_{\bm{k}_e'}^{(-)} ( \bm{r} ) &=&  \frac{1}{\sqrt{V_e}} e^{\frac{\pi}{2 k_e'}} \Gamma \bigg(1+\frac{i}{k_e'}\bigg) e^{i \bm{k}_e' \cdot \bm{ r } } \nonumber \\
&& \times ~ F \bigg(-\frac{i}{k_e'}, 1, -i (k_e' r + \bm{k}_e' \cdot \bm{ r } ) \bigg), 
\label{final_state_e_2}
\end{eqnarray} 
where $\bm{ k}_e'$ is the asymptotic momentum of the outgoing electron with the corresponding kinetic energy $\varepsilon_{{k}'_e} = \bm{ k }_e'^2/2$. The state (\ref{final_state_e_2}) takes into account the influence of the Coulomb interaction between the outgoing electron and the bound positron on each of these two particles. By dropping the (weak) interaction between the electron and the neutral core of 
$\bar{\rm{H}}^+$,   
we obtain 
\begin{eqnarray}
&&\bigg( \hat{H} - i \frac{\partial}{\partial t} \bigg) \chi_f(t)   = \bigg( \bigg[ \varepsilon_{k_e'} +  \frac{1}{r} \bigg] \phi_b (\bm{r}_p) \varphi_{\bm{k}_e'}^{(-)} ( \bm{r} )  \nonumber \\
&& \qquad \quad - \big( \hat{\bm{p}}_{\bm{r}_p} \phi_b (\bm{r}) \big) \cdot \big( \hat{\bm{p}}_{\bm{r}_e} \varphi_{\bm{k}_e'}^{(-)} ( \bm{r} )  \big)  \bigg) \, e^{-i (\varepsilon_{g}+\varepsilon_{k_e'}) t},  \nonumber \\
\label{ep_interaction_2}
\end{eqnarray}
where $\hat{\bm{p}}_{\bm{r}_p}$ ($\hat{\bm{p}}_{\bm{r}_e}$) 
is the momentum operator for the positron (electron). 

Inserting \eqref{initial_state_1} and \eqref{ep_interaction_2} into the amplitude \eqref{amplitude_exact} and calculating the resulting integrals, the transition amplitude for 3BAe is obtained to be given by
\begin{eqnarray}
a_{3BAe} &=& \frac{16 \pi^3 N (\beta^2-\alpha^2)}{i \sqrt{V_p} V_e (\alpha^2+P^2) (\beta^2+P^2)} e^{-\frac{\pi}{2 \kappa}} \Gamma \bigg(1-\frac{i}{\kappa} \bigg) \nonumber \\
&& e^{\frac{\pi}{2 k_e'}} \Gamma \bigg(1-\frac{i}{k_e'} \bigg) Q_{pe} \delta ( \Delta ),
\label{amplitude_result_2}
\end{eqnarray}
where
\begin{eqnarray}
Q_{pe} &=& \frac{1}{\tilde{\alpha}} \bigg( \frac{\tilde{\alpha}}{\tilde{\gamma}} \bigg)^{\frac{i}{\kappa}} \bigg( \frac{\tilde{\gamma}+\tilde{\delta}}{\tilde{\gamma}} \bigg)^{-\frac{i}{k_e'}} \bigg[ \Lambda_1 F \bigg(1-\frac{i}{\kappa}, \frac{i}{k_e'}, 1, z \bigg) \nonumber \\
&& - \Lambda_2 F \bigg(2-\frac{i}{\kappa}, \frac{i}{k_e'}+1, 2, z \bigg) \bigg].
\label{K}
\end{eqnarray}
In \eqref{K}, $z= \frac{\tilde{\alpha} \tilde{\delta} - \tilde{\beta} \tilde{\gamma}}{ \tilde{\alpha} ( \tilde{\gamma} + \tilde{\delta} ) }$, $\tilde{\alpha} = (q^2 + \gamma^2)/2$, $\tilde{\beta} = \bm{k}_e' \cdot \bm{q} - i \gamma k_e'$, $\tilde{\gamma} = \bm{\kappa} \cdot \bm{q} + i \gamma \kappa - \tilde{\alpha}$, $\tilde{\delta} = \kappa k_e' + \bm{\kappa} \cdot \bm{k}_e' - \tilde{\beta}$, $\bm{q} = \bm{k}_e - \bm{k}_e'$ and $F(a,b,c,z)$ is the hypergeometric function \cite{Abramowitz}. 
The quantities $\Lambda_1 (\bm{k}_p, \bm{k}_e, \bm{k}_e')$ and $\Lambda_2 (\bm{k}_p, \bm{k}_e, \bm{k}_e')$, which are quite cumbersome, are given in the Appendix. 

\vspace{0.25cm}  

The transition rate for the 3BAe process is obtained by calculating the quantity 
\begin{eqnarray}
R_{3BAe} &=& \frac{V_e}{(2 \pi)^3} \int d^3 \bm{ k}_e' \frac{| a_{3BAe} |^2}{T},
\label{rate}
\end{eqnarray}
where the integration runs over the momentum 
$\bm{ k}_e'$ of the outgoing electron and the time duration $T$ is of the order of the interaction time. 

Next, we insert the amplitude \eqref{amplitude_result_2} obtained into \eqref{rate} and express $V_p$ and $V_e$ by the corresponding number densities $n_p$ and $n_e$ of positrons and electrons according to $V_p = 1/n_p$ and $V_e=1/n_e$. Subsequently, the $\bar{\rm{H}}^+$ formation rate $R_{3BAe}$ per unit of time (per anithydrogen $\bar{\rm{H}}$) for 3BAe reads 
\begin{eqnarray}
R_{3BAe} &=& \frac{16 \pi^2 n_p n_e N^2  (\beta^2-\alpha^2)^2}{ (\alpha^2+P^2)^2 (\beta^2+P^2)^2} e^{-\frac{2 \pi}{\kappa}}  \, G(\kappa) \, G(k_c') k_c' \nonumber \\
&&\times \int d\Omega_{\bm{ k}_e'} 
(| Q_{pe} |^2)_{k_e' = k_c'}.
\label{rate_result_2}
\end{eqnarray}
Here
\begin{eqnarray}
G(k) & = & e^{\frac{\pi}{k}} \bigg| \Gamma \bigg(1-\frac{i}{k} \bigg) \bigg|^2 
 =  \frac{ 2 \pi/k }{ 1 - \exp( - 2 \pi/k   )    } 
\label{Gamow_factors}
\end{eqnarray}
is the Gamow factor for a particle with absolute momentum $k$ moving in an attractive Coulomb field (see, e.g. \cite{Landau}) and $k_c' = \sqrt{k_p^2+k_e^2- 2 \varepsilon_g}$. 

In what follows, we consider antihydrogen embedded in a gas of low energy ($\sim$ meV) electrons which is bombarded by a beam of positrons. The positron beam propagates in a fixed direction and its (sharp) energy shall vary in a relatively broad range from $\sim$ sub-meV up to $\sim$ eV. In such a situation, the formation rate of $\bar{\rm{H}}^+$ ions can be obtained by averaging   
\eqref{rate_result_2}  over the absolute value $k_2$ of the incident electron momentum using a Maxwell-Boltzmann distribution. In addition, we account for all relative orientations between the incident positron and incident electron momenta by calculating the average over the solid angle $\Omega_{\bm{ k_2}}$ of the incident electron momentum while the incident positron direction is fixed. 

Consequently, the averaged formation rate $\langle R_{3BAe} \rangle$ can be calculated according to 
\begin{eqnarray}
\langle R_{3BAe} \rangle &=& \frac{1}{4 \pi} \int d\Omega_{\bm{ k}_e} \int_{0}^{\infty} d k_e w (k_e) R_{3BAe},
\label{rate_averaged}
\end{eqnarray}
where $w$ is the Maxwell-Boltzmann distribution for the incident electron. The quantity \eqref{rate_averaged} now only depends on the absolute momentum $k_p$ of the incident positron.  

\subsection{ Three-body attachment with an "assisting" positron (3BAp)}

Let us briefly consider three-body attachment, $\rm{e}^+ + \rm{e}^+ + \bar{\rm{H}} \rightarrow \rm{e}^+ + \bar{\rm{H}}^+$, in which the incident electron involved in the 3BAe process is replaced by a second incident positron. Here, one positron is attached to $\bar{\rm{H}}$ -- driven by the positron-positron interaction -- whereas the other positron takes the energy excess.

Our calculations show that the formation rate for 3BAp is many orders of magnitude smaller compared to the rate for 3BAe. This result can be explained by the fact 
that an attraction between the incident positron and electron in the 3BAe mechanism is replaced by a repulsion 
between two incident positrons in the 3BAp which does not allow them to come close to each other weakening the attachment reaction.  

It is worth mentioning, however, that the corresponding reaction resulting in the production of an antihydrogen atom,  $\rm{e}^+ + \rm{e}^+ + \bar{\rm{p}} \rightarrow \rm{e}^+ + \bar{\rm{H}}$, can be quite efficient and, indeed, is used for the production of antihydrogen in laboratories. Compared to the formation of $\bar{\rm{H}}^+$ via 3BAp, here, the mutual repulsion of the incident positrons is balanced by their attraction to the antiproton 
(which is now not screened by the bound (anti)atomic positron).  
 
In Section III we shall consider our numerical results for 
the formation of $\bar{\rm{H}}^+$ 
via the 3BAe mechanism with the corresponding formation 
via the two radiative mechanisms: spontaneous radiative and two-center dileptonic attachments. Therefore, for convenience, we very briefly discuss them in the next subsection.

\subsection{ Radiative attachment mechanisms } 

\subsubsection{ Spontaneous radiative attachment } 
 
Spontaneous radiative attachment (SRA) of an electron to atomic hydrogen via emission of a photon has been studied in detail in the past (see, e.g., \cite{Smirnov, Janev, McLaughlin, Keating_1, Keating_2} and references therein) and the corresponding results can be straightforwardly applied also for positron attachment to antihydrogen \cite{Keating_1, Keating_2}: $\rm{e}^+ + \bar{\rm{H}} \rightarrow \bar{\rm{H}}^+ + \hbar \omega_k$ with $\omega_k$ the angular frequency of the emitted photon. To be consistent with the calculations for three-body attachment with an assisting electron in section II A., we have considered the $\bar{\rm{H}}^+$ formation via spontaneous radiative attachment using the same wavefunction \eqref{ion_bound_state_p} for the bound state of the ion. The corresponding formation rate $R_{SRA}$ of $\bar{\rm{H}}^+$ ions per unit of time (per $\bar{\rm{H}}$) for SRA reads
\begin{eqnarray} 
R_{SRA} &=& \frac{ 64 \pi^2 n_p N^2 (\beta^2-\alpha^2)^2 k_p^2 ( \varepsilon_{k_p} - \varepsilon_g ) }{ 3 \, c \, \, (\alpha^2 + k_p^2)^2 (\beta^2 + k_p^2)^2  }, 
\label{SRA_rate}
\end{eqnarray}
where $c$ is the speed of light.

\subsubsection{ Two-center dileptonic attachment }

In \cite{Jacob_1} we have considered two-center dileptonic attachment (2CDA) of a positron to $\bar{\rm{H}}$ which may occur when positrons and antihydrogen move in a (dilute) gas of atomic species B: $\rm{e}^+ + \bar{\rm{H}} + B \rightarrow \bar{\rm{H}}^+ + B^* \rightarrow \bar{\rm{H}}^+ + B + \hbar \omega_k$. If the energy release $\omega_A = \varepsilon_{k_p} - \varepsilon_g$ in the positron attachment is close to an excitation energy $\omega_B$ of a dipole transition in B, then the energy excess can be efficiently transferred, via the two-center positron-electron interaction, to B which, in consequence, undergoes a transition into an excited state. Subsequently, atom B radiatively decays to its initial (ground) state and the two-center system becomes stable which means that the $\bar{\rm{H}}^+$ ion has been formed. 

From the consideration in \cite{Jacob_1}, we obtain the $\bar{\rm{H}}^+$ formation rate $R_{2CDA}$ per unit of time (per $\bar{\rm{H}}$) for 2CDA which is given by
\begin{eqnarray}
 R_{2CDA} &=& R_{SRA} \times \frac{ 9 \pi }{ 4 } \frac{ n_B }{ v \, b_{min}^2 } \frac{ c^6 \Gamma_r^B }{ \omega_A^3 \omega_B^3 } 
\eta^2 \bigg\{ \sin^2\vartheta_{\bm{k}_p} K_1^2(\eta) \nonumber \\
&&+ \big(1 + \cos^2\vartheta_{\bm{k}_p}) 
\eta K_0(\eta) K_1 (\eta \big) \bigg\}.
\label{2CDA_rate}
\end{eqnarray}
Here, $n_B$ is the density of atoms B, $b_{min}$ is the minimum impact parameter of the $\bar{\rm{H}} - \rm{B} $ collisions and $\Gamma_r^{B}$ is the radiative width of the excited state of atom B. Further, $\eta = | \omega_A - \omega_B | b_{min}/v$, where $v$ is the velocity of $\bar{\rm{H}}$ with respect to $\rm{B} $, and $\vartheta_{\bm{k}_p}$ is the incident positron angle (counted from the collision velocity $v$). Besides, $K_j(x)$ ($j = 0,1$) are the modified Bessel functions \cite{Abramowitz}. For the derivation of the rate \eqref{2CDA_rate}, we only have taken into account contributions to 2CDA from distant collisions where $b_{min} \gg 1$ a.u. \cite{Jacob_1, Jacob_2}. Thus, the result \eqref{2CDA_rate} represent a lower boundary for the 2CDA formation rate. 

2CDA is a resonant process which relies on an energy transfer resonant to a transition energy in atom B. However, the relative motion of $\bar{\rm{H}}$ and B leads to uncertainty in positron and electron transition energies (as they are viewed by the collision partner), effectively broadening them. Therefore, the efficiency of the two-center attachment is limited to low velocity collisions, in which the velocity $v$ of $\bar{\rm{H}}$ with respect to B is much less than 1 a.u. \cite{Jacob_1, Jacob_2}.

\section{Results and discussion} 

\begin{figure}[h!]
\centering
\includegraphics[width=8.5cm]{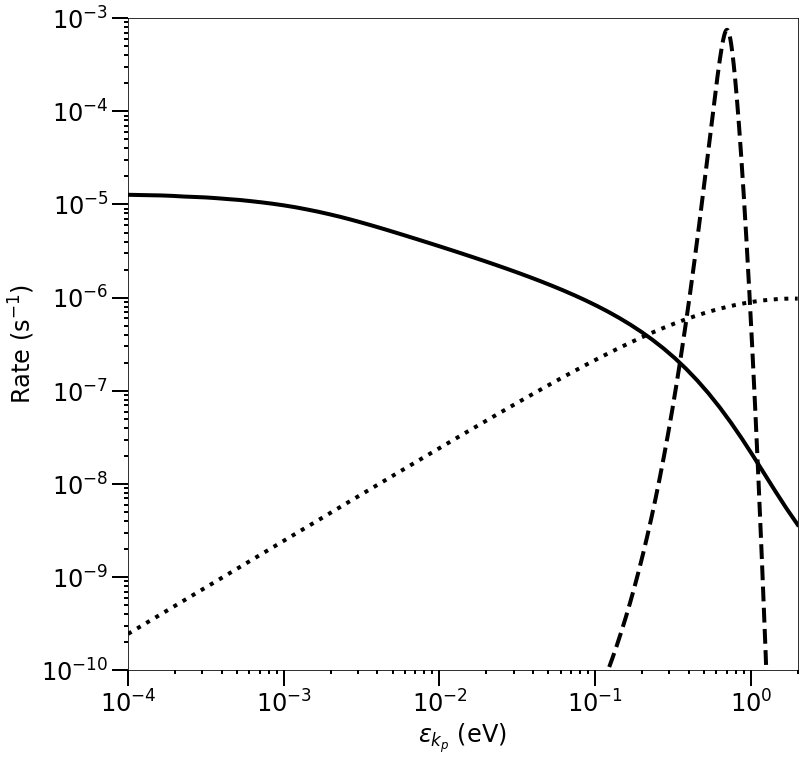}
\caption{The $\bar{\rm{H}}^+$ formation rate 
(per $\bar{\rm{H}}$) as a function of the incident positron energy $\varepsilon_{k_p}$ for three-body attachment with an assisting electron (solid curve). In addition, we illustrate the corresponding rates for spontaneous radiative attachment (dotted curve) and two-center dileptonic attachment (dashed curve). See text for further details.}
\label{fig:figure2}
\end{figure}

In Fig. \ref{fig:figure2} we show the 
$\bar{\rm{H}}^+$ formation rate \eqref{rate_averaged} via the 3BAe mechanism (the solid curve) as a function of the energy of the incident positron. Besides, for a comparison, 
in this figure we also display the formation rates \eqref{SRA_rate} and \eqref{2CDA_rate} for spontaneous radiative attachment 
(the dotted curve) and two-center dileptonic attachment 
(the dashed curve).    

For evaluating the 3BAe rate \eqref{rate_averaged}, we average over the absolute value of the incident electron momentum by using a Maxwell-Boltzmann distrubution with an average thermal energy of 1 meV ($\approx 11.6$ K), which is in the range of typical energies for electrons/positrons in the cryogenic environment of $\bar{\rm{H}}$/$\bar{\rm{H}}^+$ experiments at CERN (see, e.g. \cite{Holzscheiter, Fajans}). The density of incident electrons is chosen as $n_e = 5 \times 10^{10}$ cm$^{-3}$ that, at the moment, is the highest possible electron density experimentally achievable in a cryogenic environment of temperatures $T \sim 10$ K \cite{Mohamed}. Further,  
we set the incident positron density to $n_p = 10^8$ cm$^{-3}$ which corresponds to the typical density of positrons in $\bar{\rm{H}}$/$\bar{\rm{H}}^+$ experiments (see, e.g. \cite{Holzscheiter, Fajans}).

Concerning 2CDA, we evaluate -- following \cite{Jacob_1} -- the rate \eqref{2CDA_rate} for the case when positrons and a beam of slow $\bar{\rm{H}}$ move in a gas of Cs atoms. Here, the positron capture by antihydrogen involves the $6 ^2\rm{S}_{1/2} \rightarrow 6 ^2\rm{P}_{3/2}$ dipole transition in Cs with $\omega_B = 1.455$ eV and $\Gamma_r^B = 2.14 \times 10^{-8}$ eV. Further, we choose $n_B = 10^{15}$ cm$^{-3}$, $b_{min} = 5$ a.u., $v=0.01$ a.u. and 
$\vartheta_{\bm{k}_p} = \pi/2$. Note that if we average \eqref{2CDA_rate} over the direction of the incident positron, 
the resulting rate becomes $\approx 34 \%$ smaller than that at a fixed $\vartheta_{\bm{k}_p} = \pi/2$. 

It can be seen in Fig. \ref{fig:figure2} that in the interval of positron energies ranging from $10^{-4}$ eV to 2 eV 
the rate $\langle R_{3BAe} \rangle$ is maximal at the smallest energy shown, weakly changes between $10^{-4}$ eV and $10^{-3}$ eV and then decreases faster and faster when the energy continues to increase. The rate $R_{SRA}$ shows a different behaviour, growing with increasing $\varepsilon_{k_p}$ and then saturating at the largest energies depicted. Further, the rate $R_{2CDA}$ has a resonant structure, reaching a maximum close to the position of the resonance at $\varepsilon_{k_p,r} = \varepsilon_g + \omega_B \approx 0.71$ eV and decreasing rapidly when departing from this point.   

We can observe in Fig. \ref{fig:figure2} that the 2CDA mechanism can only be competitive close to the resonance energy at $\varepsilon_{k_p,r} \approx 0.71$ eV. In particular, exactly 
on the resonance, it dominates the SRA by a factor $\approx 9.3 \times 10^2$ and 3BAe by a factor $\approx 4.9 \times 10^3$. However, at (much) smaller positron energies $\varepsilon_{k_p} \lesssim 10^{-2}$ eV, which are most favorable for 3BAe, 
the rate $\langle R_{3BAe} \rangle$ is orders of magnitude larger than the rate $R_{SRA}$ and even many more orders of magnitude larger then the rate $R_{2CDA}$ (since the positron energy is far away from the two-center resonance).
Moreover, 3BAe remains stronger than SRA and 2CDA up to energies $\varepsilon_{k_p} \sim 0.1$ eV. 

The 3BAe and 2CDA mechanisms involve the interaction between matter and antimatter. Thus, the question naturally arises whether other processes, which will be present in such an environment (in particular, annihilation), would not effectively terminate them. 

With respect to the 2CDA mechanism, this point was discussed 
in some detail in \cite{Jacob_1} where it was shown that other processes do not have a strong impact on the efficiency of the 2CDA. 

Concerning 3BAe, the positron-electron annihilation is the only process which might influence the efficiency of this mechanism. The annihilation can occur either in a free electron-positron pair or in a (sub)system of an electron 
and a bound positron.   

The probability 
for annihilation of a free positron-electron pair is quite low. In particular, at an energy of $1$ meV for the relative 
positron-electron motion, we obtain for the annihilation cross section $4 \times 10^{-21}$ cm$^2$ \cite{Comment}. 
The corresponding mean free path of positrons in a gas of electrons is $5 \times 10^7$ m (at an electron density $n_e = 5 \times 10^{10}$ cm$^{-3}$). Consequently, this process is not expected to have any noticeable impact on the efficiency of the 3BAe mechanism. In principle, a free electron-positron pair can emit a photon and form a positronium which would eventually annihilate. However, at low energies the cross section for this process is very small and it can be neglected.  

The probability 
for annihilation of a free electron with a positron which is bound in  $\bar{\rm{H}}$  is also very low. Assuming that the relative velocity between the electron and positron is of the order of $ 1 $ a.u. we get for the annihilation cross section $ \sim 10^{-23}$ cm$^2$ which is even lower than for a free pair.  

The situation could change if the electron would be able to capture a positron from $\bar{\rm{H}}$ forming a positronium or if there would exist bound or long-lived resonance states in the $e^- - \bar{\rm{H}}$ system. However, at the energies of interest, the formation of a positronium is energetically forbidden. Besides, since 
it is known that the $e^+ - \rm{H}$ system does not have stable bound states and that at low positron energies resonances in this system are also absent (see e.g. \cite{resonances_in_pos_on_H}), we may conclude that all such states for 
the $e^- - \bar{\rm{H}}$ system are absent as well.

\section{Conclusion}

The formation of positive ions of antihydrogen via three-body attachment with an assisting electron, $\rm{e}^- + \rm{e}^+ + \bar{\rm{H}} \rightarrow \rm{e}^- + \bar{\rm{H}}^+$, was considered. 
This three-body mechanism was compared with the spontaneous radiative attachment of a positron to antihydrogen and the two-center dileptonic attachment in which a positron is captured by antihydrogen in collisions with cesium atoms.   

Our results show that with a realistic choice of 
positron and electron densities 
($\sim 10^8$ cm$^{-3}$ and $\sim 10^{10}$ cm$^{-3}$, respectively) 
the 3BAe can be by far the dominant attachment mechanism at very low ($\lesssim 10^{-2}$ eV) positron/electron energies. At the parameters considered 
its efficiency is essentially not influenced by annihilation and exceeds by  
orders of magnitude that of the corresponding 
three-body reaction $\rm{e}^+ + \rm{e}^+ + \bar{\rm{H}}$ where the role of an assisting particle 
is played by a positron.   

One should add that we have considered the three-body attachment using a rather simple approach which is not expected to yield very accurate results. Nevertheless, at low energies the dominance of the three-body mechanism over the other attachment processes 
was found to be so extremely strong that this finding would hardly be changed in a more elaborate treatment. 

\section*{Acknowledgement} 

We acknowledge the support from the Deutsche Forschungsgemeinschaft 
(DFG, German Research Foundation) under Grant No 349581371 
(MU 3149/4-1 and VO 1278/4-1). 

\section*{Appendix}

Here we present 
the quantities $\Lambda_1 (\bm{k}_p, \bm{k}_e, \bm{k}_e')$ and $\Lambda_2 (\bm{k}_p, \bm{k}_e, \bm{k}_e')$ 
which enter expressions \eqref{amplitude_result_2} and \eqref{K}:
\begin{eqnarray}
 \Lambda_1 &=& 1 - \varepsilon_{k_e'} A_1 - i k_e' \bm{P} \bigg( i \frac{\bm{k}_e'}{k_e'} A_1 + \bm{A}_3 + \bm{A}_5 \bigg), \nonumber \\
 \Lambda_2 &=& \varepsilon_{k_e'} A_2 + i k_e' \bm{P} \bigg( i \frac{\bm{k}_e'}{k_e'} A_2 + \bm{A}_4 + \bm{A}_6 \bigg)
\end{eqnarray}
with
\begin{eqnarray}
 A_1 &=& \frac{i}{\kappa} \frac{ \gamma (\tilde{\alpha} + \tilde{\gamma}) - i \kappa \tilde{\alpha} }{ \tilde{\alpha} \tilde{\gamma} } - \frac{\gamma}{\tilde{\alpha}} - \frac{i}{k_e'} \frac{ \gamma \tilde{\delta} + i ( k_e' \tilde{\gamma} - \kappa \tilde{\delta}) }{ (\tilde{\gamma}+\tilde{\delta}) \tilde{\gamma} }, \nonumber \\
 A_2 &=& \frac{i}{k_e'} \frac{1 - i/\kappa}{ \tilde{\alpha}^2 ( \tilde{\gamma}+\tilde{\delta})^2 } \nonumber \\
&& \{ (\gamma (\tilde{\delta}+\tilde{\beta}) + i k_e' (\tilde{\alpha} + \tilde{\gamma}) - i \kappa \tilde{\beta} ) \tilde{\alpha} ( \tilde{\gamma} + \tilde{\delta}) \nonumber \\
&& - (\tilde{\alpha} \tilde{\delta} - \tilde{\beta} \tilde{\gamma}) (\gamma (\tilde{\gamma} + \tilde{\delta} - \tilde{\alpha}) + i \tilde{\alpha} (\kappa + k_e')) \}, \nonumber \\
 \bm{A}_3 &=& \frac{\bm{q}}{\tilde{\alpha}} + \frac{i}{\kappa} \frac{ \tilde{\alpha} \bm{\kappa} - (\tilde{\alpha}+\tilde{\gamma}) \bm{q} }{ \tilde{\alpha} \tilde{\gamma} } \nonumber \\
&& - \frac{i}{k_e'} \frac{ (1+ \kappa/k_e' + \gamma i/k_e') \tilde{\gamma} \bm{k}_e' - (\tilde{\gamma}+\tilde{\delta}) (\bm{q} - \bm{\kappa}) }{ (\tilde{\gamma}+\tilde{\delta}) \tilde{\gamma} }, \nonumber \\
\bm{A}_4 &=& \frac{i}{k_e'} \frac{1 - i/\kappa}{ \tilde{\alpha}^2 ( \tilde{\gamma}+\tilde{\delta})^2 } \nonumber \\
&& \{ ( - \tilde{\delta} \bm{q} + \tilde{\alpha} (\bm{\kappa} - \bm{q} + (1+\kappa/k_e'+\gamma i/k_e') \bm{k}_e') \nonumber \\
&& - \tilde{\gamma} (\bm{q} - (1+\gamma i/k_e') \bm{k}_e') - \tilde{\beta} (\bm{q}- \bm{\kappa}) ) \tilde{\alpha} (\tilde{\gamma}+\tilde{\delta}) \nonumber \\
&& - ( \tilde{\alpha} \tilde{\delta} - \tilde{\beta} \tilde{\gamma} ) ( - (\tilde{\gamma}+\tilde{\delta}) \bm{q} + \tilde{\alpha} ( 1+\kappa/k_e' + \gamma i/k_e') \bm{k}_e' ) \}, \nonumber \\
 \bm{A}_5 &=& \frac{i}{\kappa} \frac{ \tilde{\gamma} \bm{q} - \tilde{\alpha} (\bm{\kappa}-\bm{q}) }{ \tilde{\alpha} \tilde{\gamma} } - \frac{\bm{q}}{\tilde{\alpha}} \nonumber \\
&& - \frac{i}{k_e'} \frac{ \tilde{\gamma} (\bm{\kappa} - \bm{q} - \bm{k}_e') - (\tilde{\gamma}+\tilde{\delta}) (\bm{\kappa}-\bm{q}) }{ (\tilde{\gamma}+\tilde{\delta}) \tilde{\gamma} }, \nonumber \\
\bm{A}_6 &=& \frac{i}{k_e'} \frac{1 - i/\kappa}{ \tilde{\alpha}^2 ( \tilde{\gamma}+\tilde{\delta})^2 } \nonumber \\
&& \{ ( \tilde{\delta} \bm{q} - (\tilde{\alpha}+\tilde{\gamma}) \bm{k}_e' - \tilde{\beta} (\bm{\kappa} - \bm{q})) \tilde{\alpha} (\tilde{\gamma}+\tilde{\delta}) \nonumber \\
&& - ( \tilde{\alpha} \tilde{\delta} - \tilde{\beta} \tilde{\gamma} ) ( ( \tilde{\gamma}+\tilde{\delta}) \bm{q} + \tilde{\alpha} (\bm{\kappa}-\bm{q}-\bm{k}_e') ) \}.
\end{eqnarray}


\end{document}